\documentclass{aa}
\usepackage{txfonts,graphicx,amssymb,subfigure}
\usepackage[normalem]{ulem}

\begin{document}
\title{The formation of regular interarm magnetic fields in spiral galaxies}

   \author{D. Moss\inst{1}\fnmsep
\thanks{Corresponding author, \email{moss@ma.man.ac.uk}}
          \and
          R. Stepanov\inst{2}
          \and
          M. Krause\inst{3}
          \and
          R. Beck\inst{3}
          \and
          D. Sokoloff\inst{4}
          }

   \offprints{D.Moss}

   \institute{School of Mathematics, University of Manchester,
Oxford Road, Manchester, M13 9PL, UK
  \and
  Institute of Continuous Media Mechanics, Koroleva, 1,614013,  Perm, Russia
   \and
  MPI f\"ur Radioastronomie, Auf dem H\"ugel 69, 53121 Bonn, Germany
               \and
    Department of Physics, Moscow State University, 119991, Moscow, Russia }

   \date{Received ..... ; accepted .....}

\abstract{Observations of regular magnetic fields in several nearby galaxies reveal magnetic arms situated between the material arms. The nature of these magnetic arms
is a topic of active debate.  Previously, we found a hint that taking into account the effects of injections of small-scale magnetic fields (e.g. generated by turbulent dynamo action)  into the large-scale galactic dynamo can result in magnetic arm formation.
}
{We now investigate the joint roles of an arm/interarm turbulent diffusivity contrast  and injections of small-scale magnetic field on
the formation of large-scale magnetic field ("magnetic arms") in the interarm region.}
{We use the relatively simple "no-$z$" model for the galactic dynamo.
This involves projection onto the
galactic equatorial plane of the azimuthal and radial magnetic field components;
 the field component
orthogonal to the galactic plane is estimated from the solenoidality condition.
 }
{We find that the addition of diffusivity gradients to the effects of magnetic field injections makes  the magnetic arms much 
more pronounced. In particular, the regular magnetic field component becomes larger in the interarm space than that within the material arms.}
{The joint action of the turbulent diffusivity contrast and small-scale magnetic field injections (with the possible participation of other effects previously suggested) appears to be a plausible explanation for the phenomenon of magnetic arms. }


\titlerunning{Regular interarm magnetic fields}
\authorrunning{Moss et al.}

\maketitle

\section{Introduction}
\label{int}

Dynamo modelling of galactic magnetic fields has a long history, being particularly
intensive over the last 30 years. We do not intend to give a
comprehensive review here, but Ruzmaikin et al. (1988) can be mentioned
as a seminal work, and comprehensive reviews were given by Beck et al. (1996),
Brandenburg (2015);  see also Beck et al. (2015) for more recent
developments.
Substantial progress has been made towards
understanding the basic mechanisms of dynamo excitation, including both
detailed direct numerical simulations in "boxes", and also some more 
detailed models for global field structure.

A conspicuous feature of some "grand design" spiral galaxies (e.g. M81, NGC6946)
is the presence in the regular (large-scale) magnetic field of prominent magnetic arms,
situated in the interarm regions
between the material arms, as delineated by regions of active
star formation. Such arms are not always well pronounced
or complete, and may include a number of filaments, as in IC~342 (Beck 2015).
The origin of magnetic arms has attracted significant attention, but so far
there is no completely satisfactory explanation of their origin. 
Early studies include
Moss (1998) and Shukurov (1998) who in the context of simple mean field
dynamo models appealed to variations in the alpha coefficient and turbulent
resistivity ($\eta$) that were modulated by the location of the material
arms;  specifically, it can be expected that $\eta$ will be enhanced by
the additional turbulence associated with star forming regions (SFRs).

Later relevant studies include Chamandy et al. (2013, 2014, 2015) who use
a sophisticated mean field dynamo model and argue that enhanced vertical
outflows within the arms regions will preferentially remove large-scale
fields from there. Small-scale helicity is also removed by the vertical flows,
preventing catastrophic quenching. Additionally, they include modulations of
 the alpha term. These models do produce some of the
observed properties, but possibly are not completely satisfactory. For
example, these models require dynamo numbers that are close to marginal
to generate magnetic arms efficiently, and the outflows 
cannot be too strong. Gradients of turbulent diffusivity are ignored
and pitch angles of the fields are rather small compared to those of typical galaxies.

Moss et al. (2013) took a somewhat different approach, modelling the effects
of SFRs in the material arms by stochastic injection of small-scale
magnetic field within the arms, supposedly the result of small-scale dynamo
action in the SFRs. Their model produces (maybe unsurprisingly) a satisfactory
enhancement of small-scale fields in the arms, but no significant enhancement
of interarm regular (large-scale) fields.

A significant omission from the last paper, for perceived technical reasons,
was an enhanced turbulent diffusivity in the arms associated with the assumed
turbulence driven by the star formation. The authors speculated that
this might be a significant omission.
The presence of such variations in the turbulent diffusivity
cannot be verified directly by observations, but given our understanding
of the physical processes operating, it appears a plausible assumption.
Accordingly, we here present similar models to those in Moss et al. (2013), but now including
the terms associated with inhomogeneities in $\eta$ in the dynamo equations
(including the diamagnetic terms), and we demonstrate that regular magnetic arms
located between the material arms are indeed produced.
We note in passing that contrasts in diffusivity between disc and
halo regions have been included in dynamo models for at least 25 years  
(see e.g. Brandenburg et al. 1992).

\begin{figure*}
\begin{center}
\begin{tabular}{ll}
\includegraphics[width=0.45\textwidth]{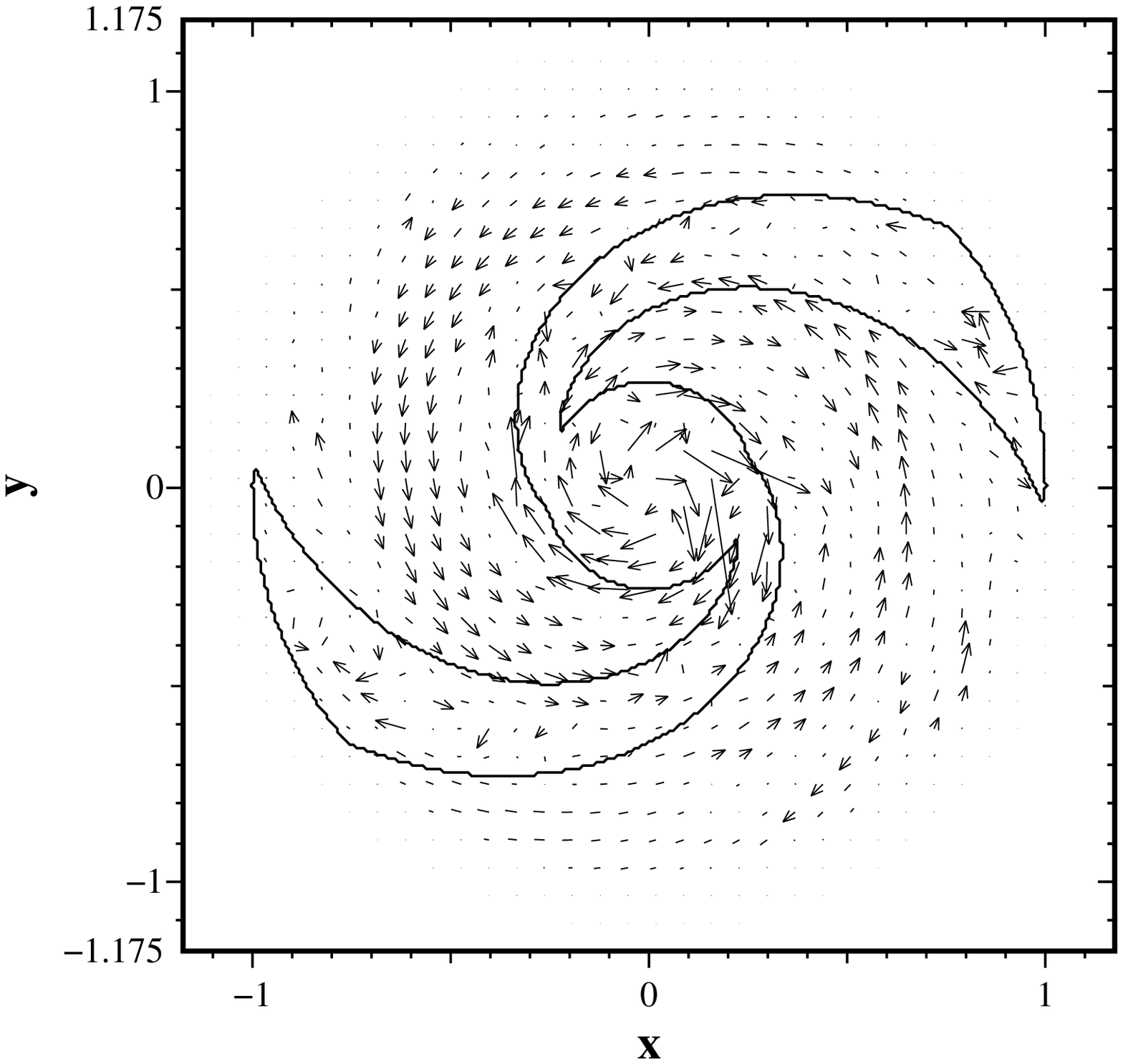} &
\includegraphics[width=0.45\textwidth]{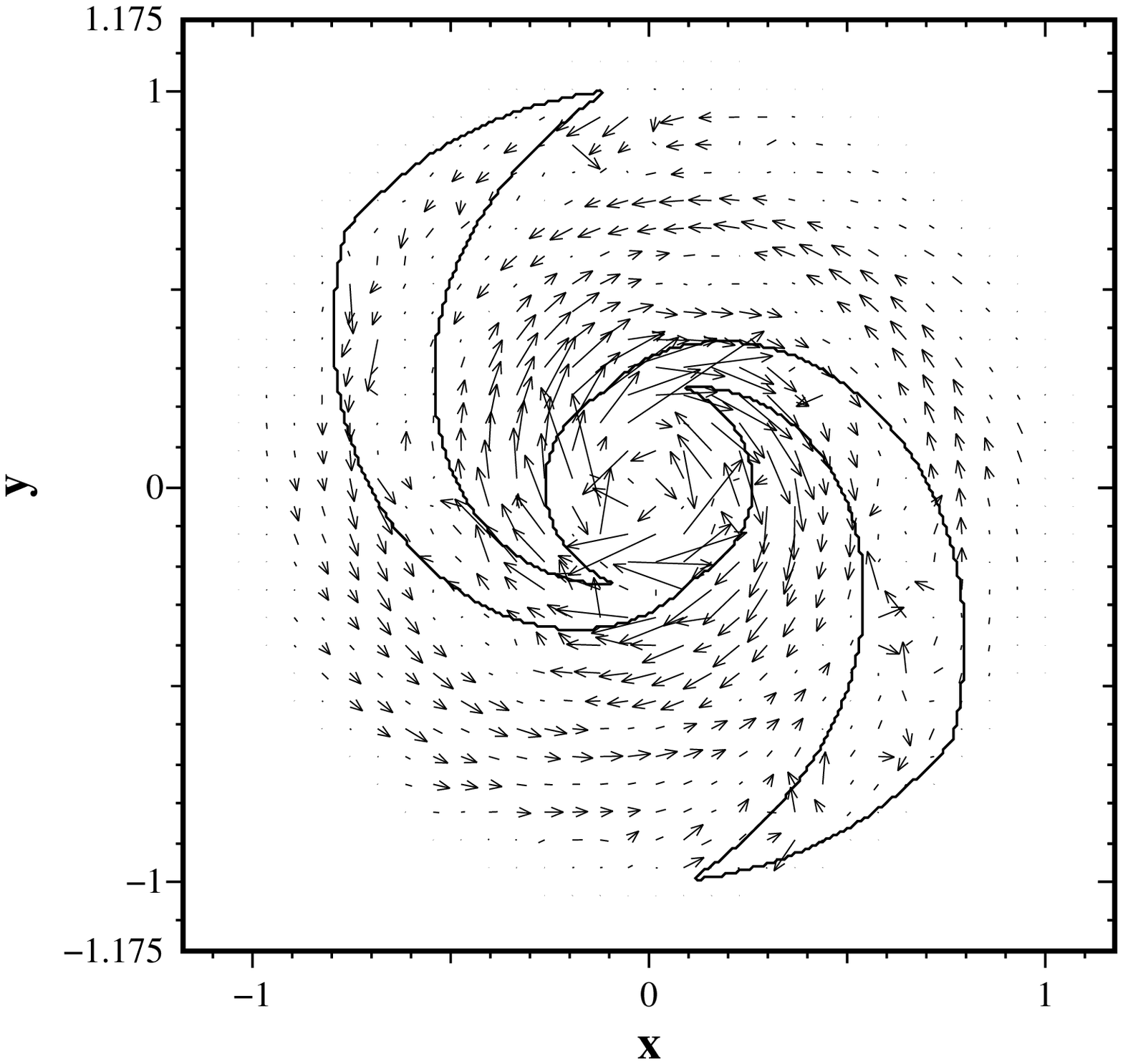} \\
\end{tabular}
\end{center}
\caption{Field vectors at dimensionless times $t\approx 3.0$ and 17.0 (2.3 and 13.3 Gyr),
for Model 2 with the diffusivity contrast parameter $\eta_{\rm a}=4$. (Here,
and in subsequent similar Figures, the vectors
give the magnetic field direction, with lengths
proportional to the magnetic field strengths, and the contours delineate the arms. The corotation radius is approximately $0.7R$.}
\label{mod182}
\end{figure*}

\section{The model}
\label{model}

\subsection{The dynamo setup}

The dynamo equation is

\begin{equation}
\frac{\partial{\bf B}}{\partial t}=\nabla\times\left(\Omega{\bf r\times B}+\alpha{\bf B}-\frac{1}{2}\nabla\eta\times {\bf B}\right) -\nabla\times(\eta\nabla\times{\bf B})
\label{dyneqn}
\end{equation}
in the standard notation.
Note the presence of gradients of diffusivity, both in the diamagnetic term and the diffusion term.
The model is basically the thin disc model ("no-$z$" approximation)
 described in Moss et al. (2012, 2013).
In these earlier papers it had not been possible to include terms
corresponding to gradients of $\eta$. The relevant part of the algorithm was
reorganized slightly, and a typo in the code corrected, after which
the code ran smoothly with gradients of $\eta$ included. (The typo did
not affect the part of the code used in earlier papers.)
 The  novel feature of the models of Moss et al. (2012, 2013) is
the continual stochastic injection of small-scale
field at discrete locations, to simulate the
effects of star forming regions in introducing small-scale field into the ISM.
In short, random fields $B_{\rm inj}=B_{\rm inj0}f(r,t)$
are added at approx $n_{\rm spot}=75$ randomly chosen discrete locations in the material arms (defined below)
with re-randomization
(i.e. changing the location of the injection sites and the
distribution of field strengths over them by choosing a new independent set of random numbers)
 at intervals $dt_{\rm inj}\approx 10$ Myr.
$n_{\rm spot}$ and $B_{\rm inj0}$ are free parameters in the model, which are regarded
as a proxies for unmodelled processes in the SFRs.
Another key point is that the seed field at time zero is random, in discrete patches, and of approximately equilibrium strength. This is envisaged
as being the result of small-scale dynamo action within very early SFRs.
(Note that the no-$z$ approximation implicitly preserves
the solenoidality condition $\nabla\cdot {\bf B}=0$ for
both the dynamo generated and injected fields.)
Full details are in Moss et al. (2012).
The disc can be flared or flat, noting that there
is currently some uncertainty as to whether galactic discs are substantially
flat or flared (cf. Lazio \& Cordes 1998); further
investigation of this point  is needed, but the results are not
sensitive to this assumption.
The HI disc of the Milky Way does flare, but it is unclear whether the
ionized gas disc does so, and the observational data for external galaxies
are inconclusive. This issue appears unimportant for our modelling --
see  Sect.~\ref{main} and also Moss et al. (2012). 

Non-dimensional time $\tau$ is measured in  units of $h^2/\eta$. When
$\eta=10^{26}$ cm$^2$ s$^{-1}$ and $h=500$ pc, this is approximately
$0.78$ Gyr. Radius $r$ is measured in units of the galactic radius $R$, taken as $10$ kpc.

\begin{figure*}
\centering
\includegraphics[width=0.32\textwidth]{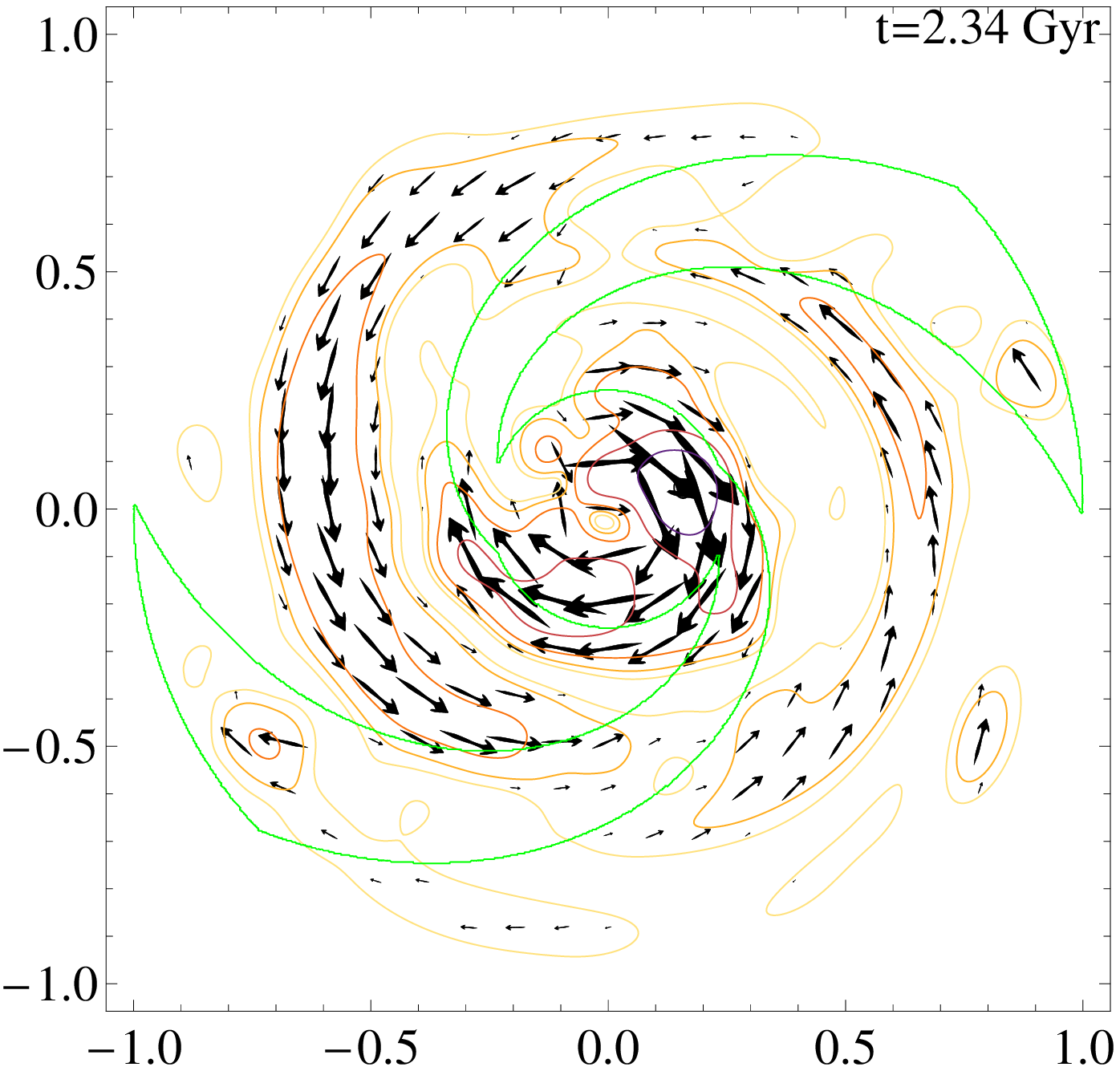}
\includegraphics[width=0.32\textwidth]{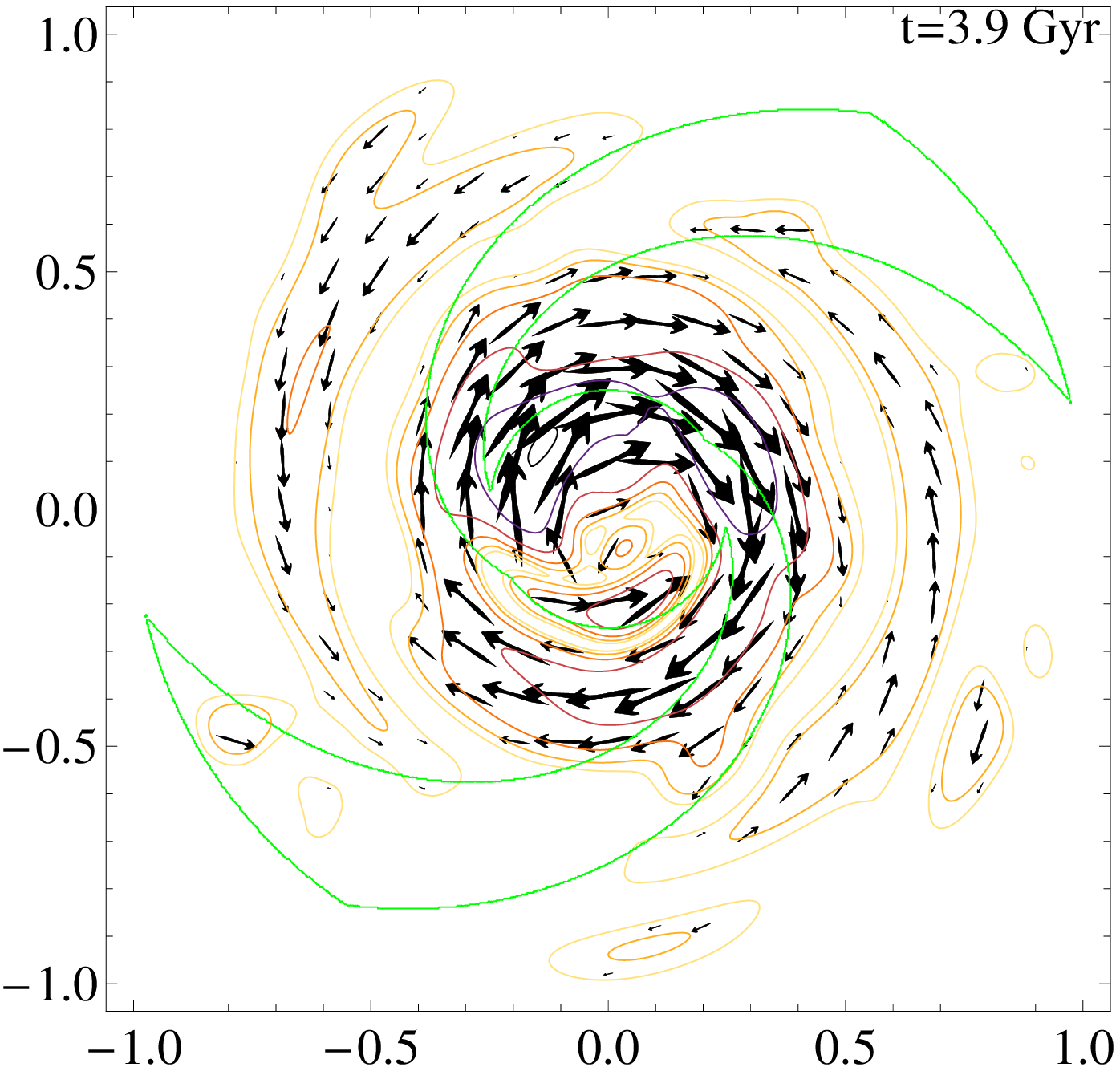}
\includegraphics[width=0.32\textwidth]{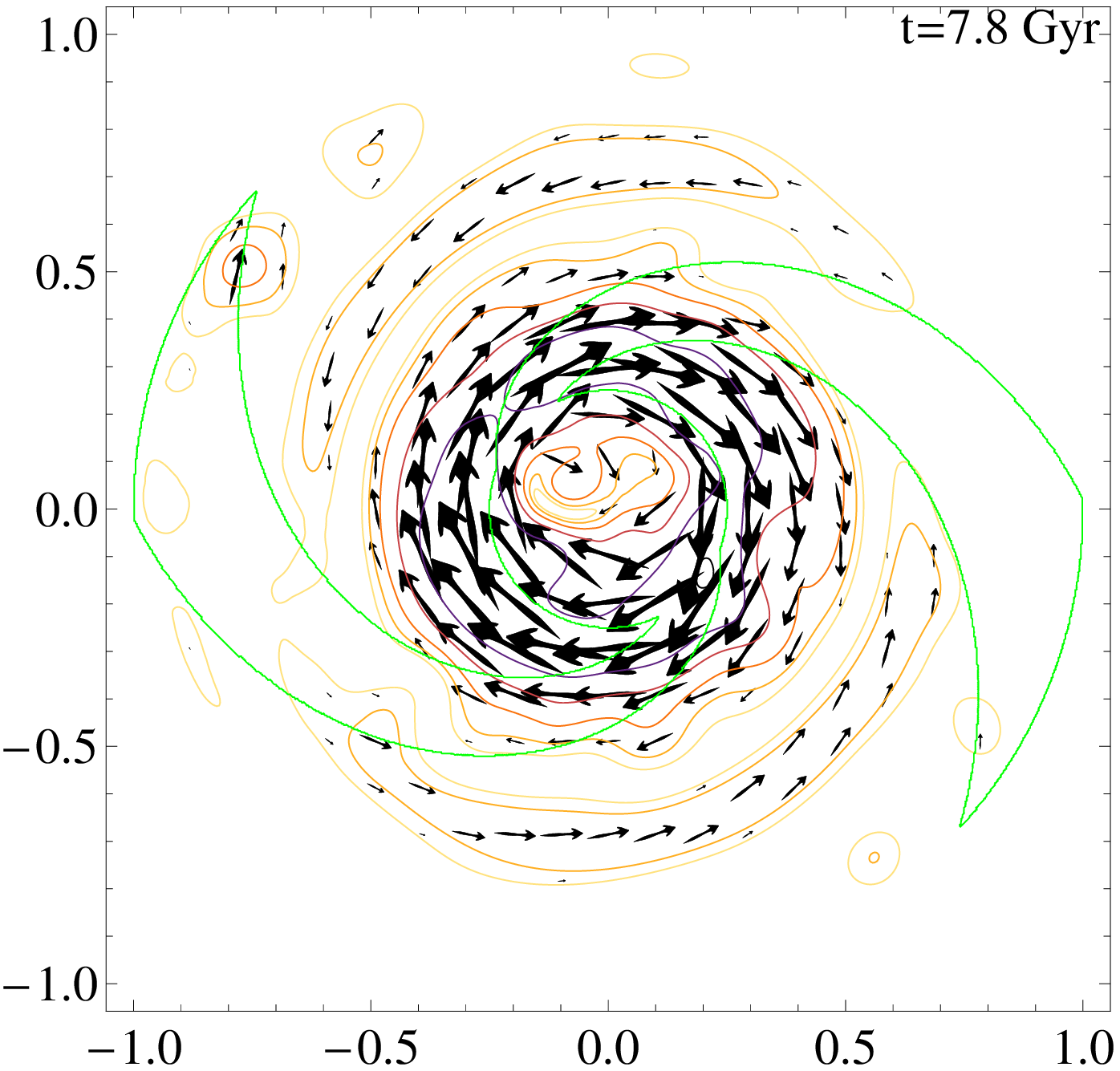}
\includegraphics[width=0.32\textwidth]{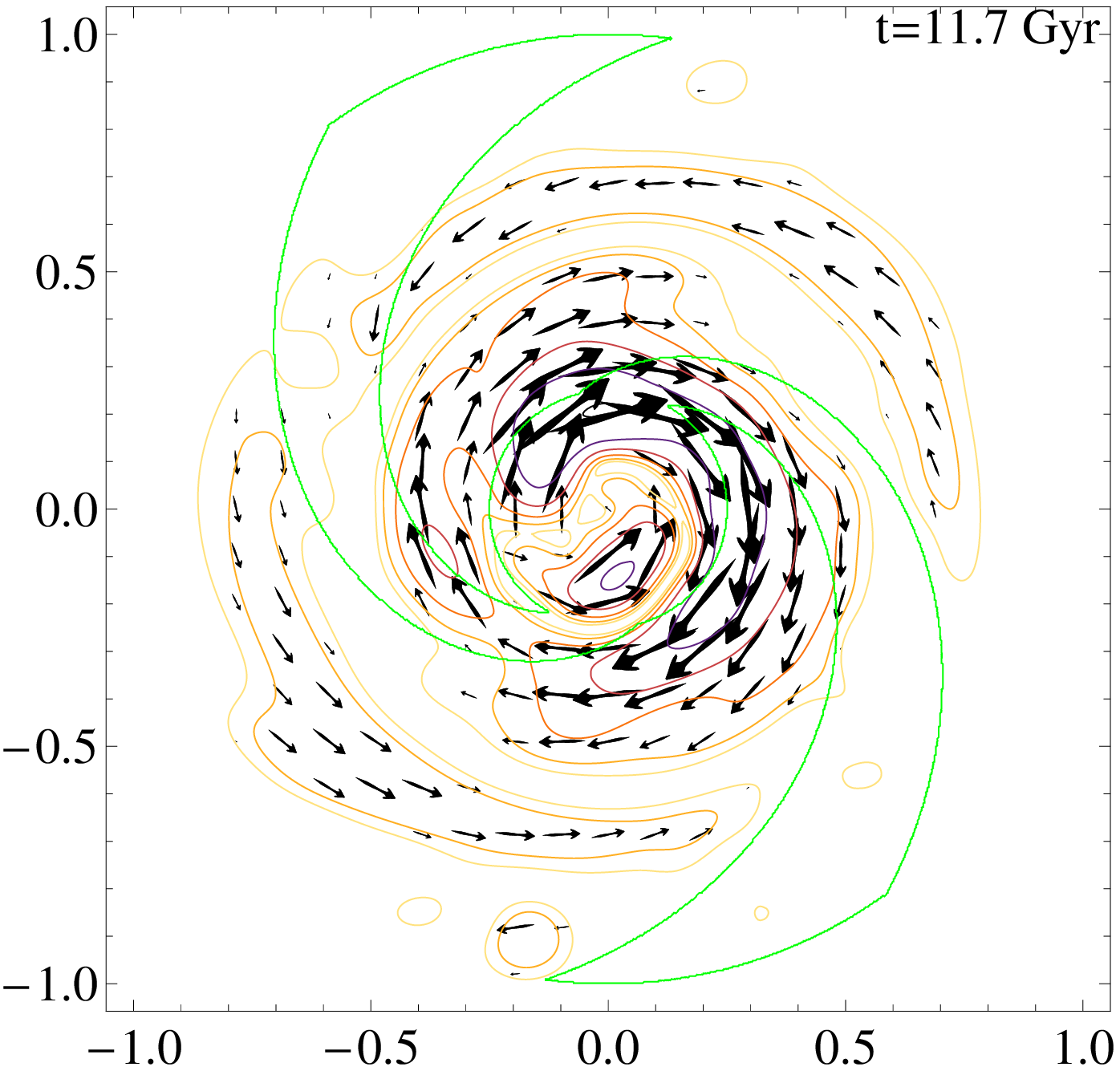}
\includegraphics[width=0.32\textwidth]{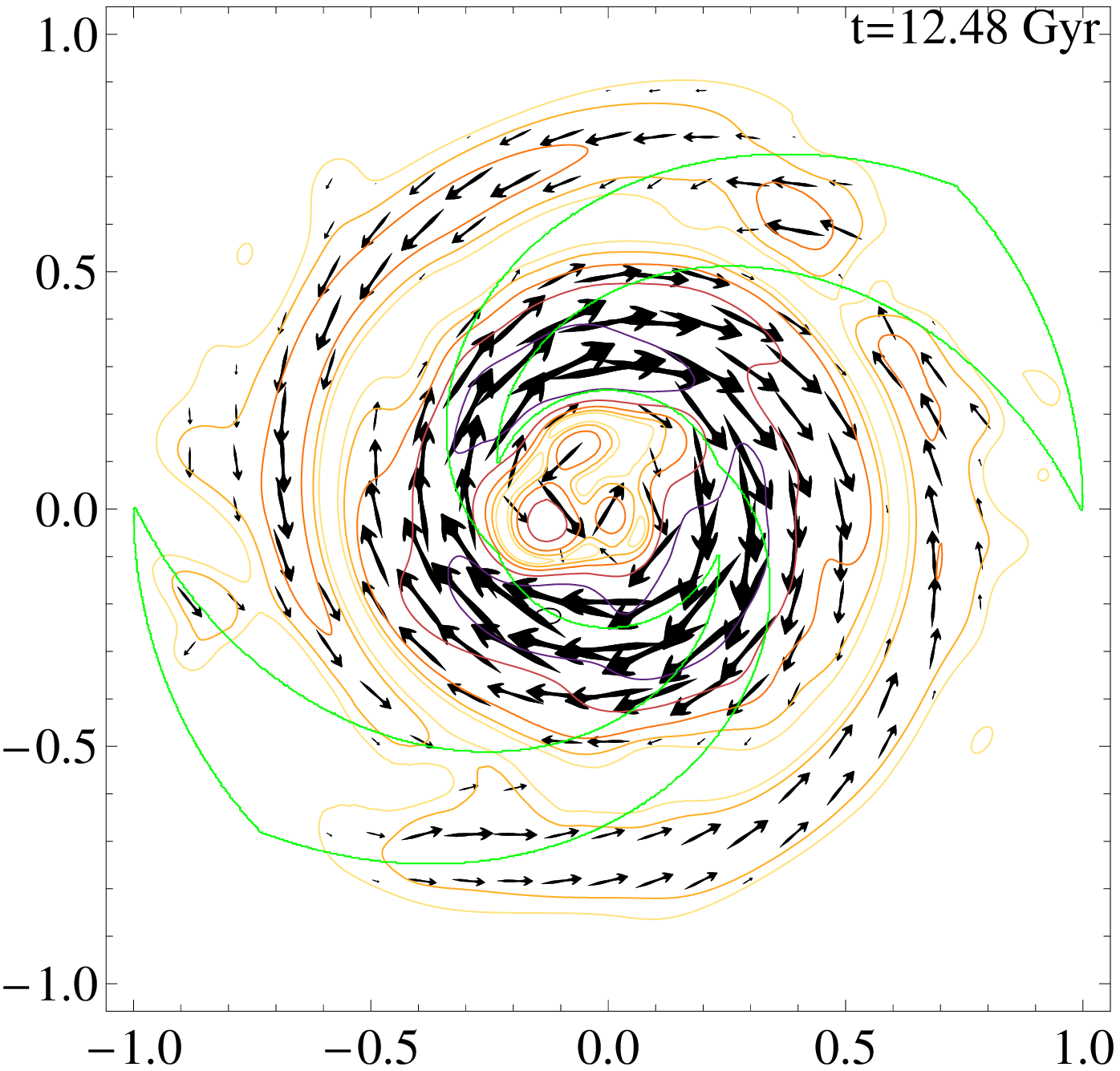}
\includegraphics[width=0.32\textwidth]{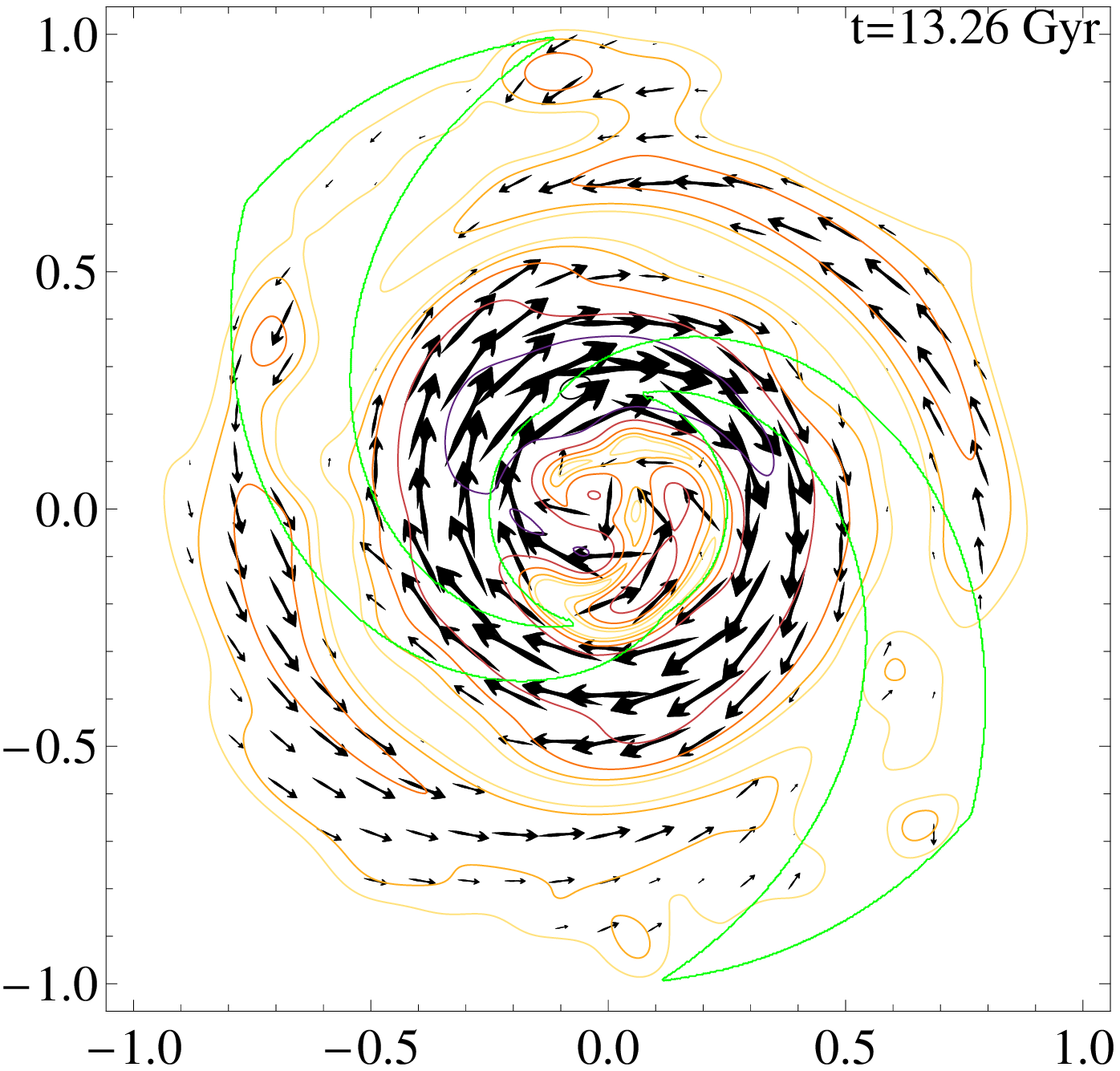}
\caption{Model 2. Intensity of the large-scale magnetic field at times
2.34, 3.9, 7.8, 11.7, 12.48 and 13.26 Gyrs (left to right, top to bottom).
Contours shows isolines of large-scale field intensity
(darker corresponds to a larger value).
The first and last panels can be compared with the panels of Fig.~\ref{mod182}. } \label{btotal_182t}
\end{figure*}

The code was implemented on a Cartesian grid with $537\times 537$ points,
equally spaced, extended to beyond the galactic radius to about
$1.17R$, i.e. over $-1.17R\la R \la 1.17R$.
In this  outer region beyond $r=R$, there is no alpha-effect and the diffusivity retains
its global background value.
This enables satisfactory treatment of the boundary conditions -- see
Moss et al. (2012).
The timestep is fixed at approximately $0.04$ Myr.

\subsection{The arm generation algorithm}
\label{arm}

We define a function $$p=\cos(0.5(2\phi-b\log(r/r_a)-2\phi_0)),$$
where $r_a, b$ are arbitrary values
and $\phi_0=\omega_p\tau$, where $\omega_p$ is the dimensionless pattern speed.

Then we put $$g=\exp(-(\frac{p}{a})^2)\eta_1(r),$$
where $\eta_1=1$, $r>0.4$, and goes smoothly to zero as $r\rightarrow 0$.
Then $$\eta=1+\eta_{10}g^m,$$
where $m$ is an arbitrary number. The diffusivity contrast is 
defined as $\Delta\eta_a=1+\eta_{10}$.

This algorithm is rather arbitrary, but gives satisfactory results
with the values $r_a=1, a=\pi/4, b=4$ (as in Moss et al. (2012)).
In the illustrative computations described below $m=4$.

The arms rotate rigidly with pattern speed $\omega_p$, chosen to give
corotation radii $0.5R\la r_{\rm corr}\la 0.7R$ -- see Table~1.

\section{Results}
\label{results}

\subsection{Main computations}
\label{main}

\begin{figure}
\includegraphics[width=0.45\textwidth]{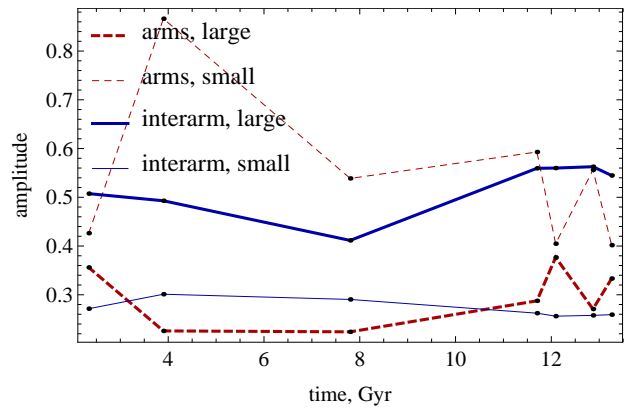}
\caption{Model 2, showing global amplitudes of large- and small-scale field in the arms
(respectively thick and thin red dashed) and large- and small-scale field in the
interarm regions (respectively thick and thin blue curves).}
\label{arm-interarm_182t}
\end{figure}

A number of models were computed, with both flat and flared discs. The
essential features of the results were the same for each class of model,
and so only results for flat discs are presented here.
Salient parameters are shown in Table~\ref{table}.
The underlying models are in general similar
to Model 75 of Moss et al. (2013), for example, but here with a flat disc.
We take $h_0=500$ pc and a slightly larger value $R_\alpha=5$ is taken to compensate for
the increased value of the diffusivity $\eta$ in the arms.
Plots of field vectors at dimensionless times $\tau=3$ and $17$ are shown in
Fig.~\ref{mod182}, with the diffusivity contrast parameter $\Delta\eta_{\rm a}=4$.
The strong visual impression is that the large-scale field {\em is} reduced
in the arms, and that there are pronounced large-scale magnetic structures
(corresponding to regular field) {\em between} the material arms.
Moreover, these effects are seen at very early times.
This impression is reinforced  (in a somewhat different representation)
by Fig.~\ref{btotal_182t} 
, showing maps of 
the large-scale field
at dimensionless times between $\tau=3$ and $17$ approximately
$2.3$ and $13.3$ Gyr.
This figure can be compared to Fig.~4 of Moss et al. (2013).
Additionally, Fig.~\ref{arm-interarm_182t} shows the global amplitudes
of large-and small-scale fields in the arms and interarm regions.
This figure can be compared to Fig.~8 of Moss et al. (2013), and
clearly demonstrates the effects of the assumed $\eta$-gradients, in that
in the interarm region the large-scale
(regular) field is now two or three times larger than in the arms.

Model~2 has corotation radius at approximately $0.7R$. We computed models with
different corotation radii, and it appears that the results are
quite insensitive to such changes.
For example, Fig.~\ref{mod183} shows field vector plots at $\tau=17$
for a model with corotation  at approximately $0.5 R$. (The anomalously large
vectors in Fig.~\ref{mod183} (and \ref{mod186}c below) are attributable to
a field injection soon before plotting.)

\begin{figure}
\includegraphics[width=0.45\textwidth]{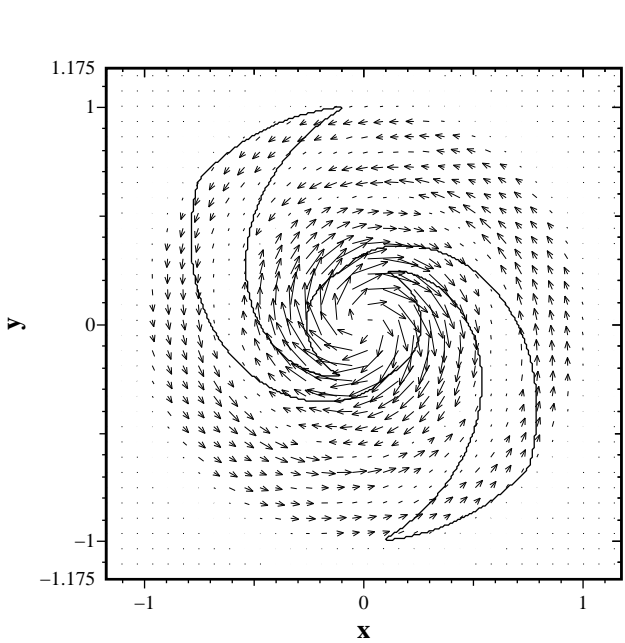}
\caption{Model 7. Computation starting from a random seed field (as Model~2),
but with no ongoing small-scale field injections. $\Delta\eta_a=4$.}
\label{mod187}
\end{figure}

\begin{figure}
\begin{center}
\includegraphics[width=0.45\textwidth]{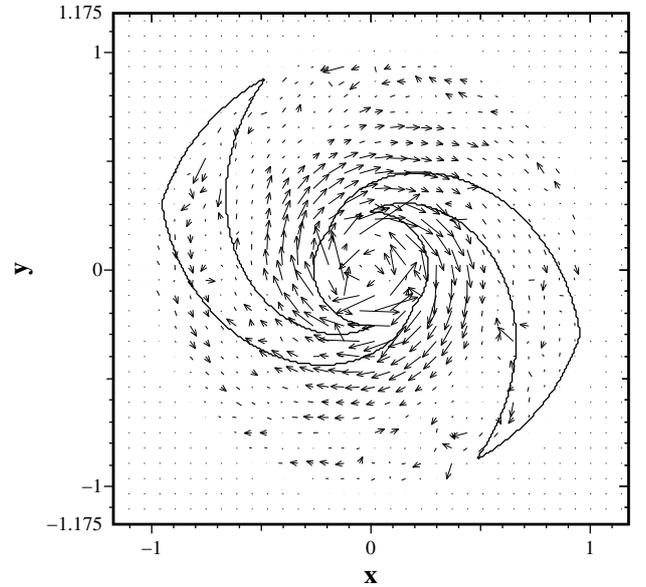} \\
\end{center}
\caption{Field vectors at dimensionless time  $\tau=17.0$ (13.3 Gyr),
for model 3 with diffusivity contrast parameter $\eta_{\rm a}=4$.
The corotation radius is approximately $0.5R$.
}
\label{mod183}
\end{figure}

\begin{figure}
\begin{center}
\includegraphics[width=0.45\textwidth]{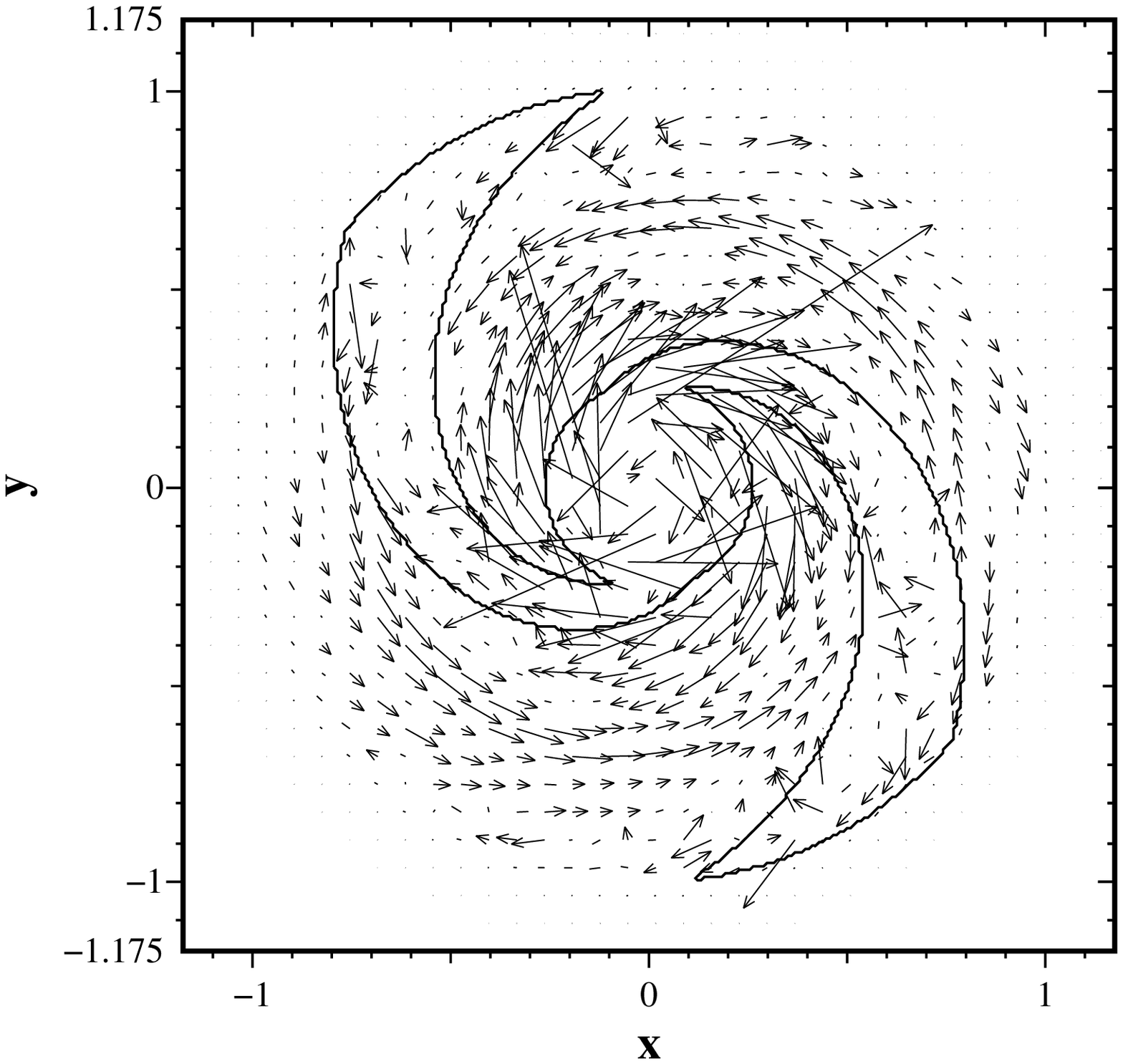} \\
\end{center}
\caption{Field vectors at dimensionless time  $\tau=17.0$ (13.3 Gyr),
for Model~8 with the larger diffusivity contrast parameter $\eta_{\rm a}=6$.
The corotation radius is approximately $0.7R$.
}
\label{mod188}
\end{figure}

The value taken for the diffusivity contrast parameter $\Delta\eta_a$ is rather arbitrary, so the computation
of Model~2 (Fig.~\ref{mod182}) was repeated with $\Delta\eta_a=6$, see Fig.~\ref{mod188}. The magnetic arms are more marked and the interarm magnetic
structures are somewhat narrower. Further increase in $\Delta\eta_a$ can give more
filamentary magnetic 'arms'.

\subsection{Field reversals}
\label{reverse}

\begin{figure}
\begin{center}
\includegraphics[width=0.45\textwidth]{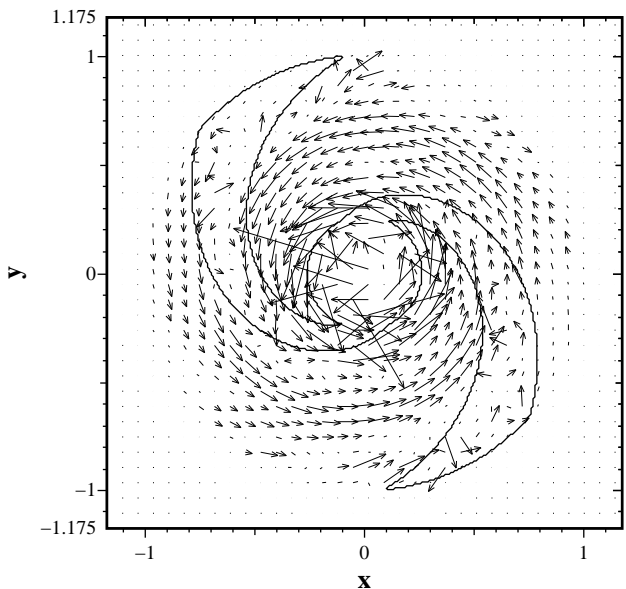} \\
\end{center}
\caption{Field vectors at dimensionless time  $\tau=17.0$ (13.3 Gyr),
for Model 9 with parameters as Model~2, but a different sequence of
random numbers defining the initial field and subsequent injections.
}
\label{mod189}
\end{figure}

Large-scale field reversals are a common feature of the models and
were also present in the models of Moss et al. (2012, 2013) see also
Poezd et al. (1993).
Moss \& Sokoloff (2013) showed that when the seed field is inhomogeneous and
of near equipartition strength, whether or not reversals appeared
in the statistically steady field (and indeed the details
of this configuration more
generally) depends quite sensitively on the details of the initial field distribution.
Several models were run with the same parameters as Model~2, but using different
sequences of the (pseudo-)random numbers that define the initial field distribution
and subsequent field injections.
The field configurations at the end of the runs were
generically similar, several with reversals but one (Fig.~\ref{mod189})
with only a very weak feature near the outer radius.

\begin{figure}
\begin{center}
\includegraphics[width=0.45\textwidth]{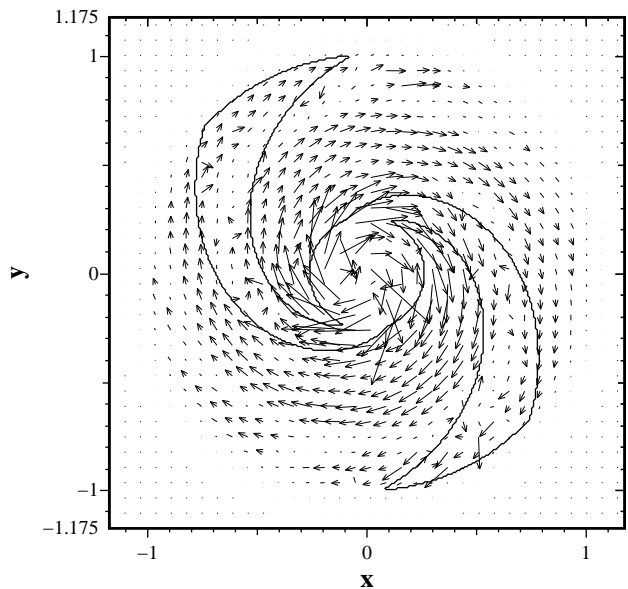} \\
\end{center}
\caption{Field vectors at dimensionless time  $\tau=17.0$ (13.3 Gyr),
for Model~19. Here the seed field was weak and smooth, and there were no
random injections until $\tau=8$.
The corotation radius is approximately $0.7R$.
}
\label{mod199}
\end{figure}

We also made an experiment in which the seed field was weak and smooth, and the model was allowed
to evolve without injections until the dynamo was saturated, with a "standard" looking smooth spiral
structure. At $\tau=8$ the injections were turned on and evolution proceeded 
as in the other cases. The field at $\tau=17$ is shown in Fig.~\ref{mod199}.
Now there are no large-scale reversals, but other features of the field are similar to those in other models. This clearly
demonstrates that the occurrence of large-scale reversals is a consequence of 
the form of the seed field and the injection history.

The models presented above all have a more or less homogeneous inner ring
of field, with a reversal between the ring and the outer arms. 
A field reversal has been observed in the Milky Way, but it is
unclear whether it is a global or local feature. Reversals have
not (so far) been detected in
external galaxies. In some ways the presence of the ring is a consequence of the way the model is set up,
as the arms are only distinctly defined outside of the central regions
(see Moss et al. 2013).

\begin{table*}
\caption{Summary of models}
\begin{center}
\begin{tabular}{llll} \hline \noalign{\smallskip}
 Model&$\Delta\eta_a$&$r_{\rm corr}$ & comment\\
\hline
\noalign{\medskip}
2&4&0.7&basic model\\
3&4&0.5&\\
7&4&0.7& no ongoing field injections\\
6&4&0.7& as Model~2, no injections between $\tau=8.5$ and 10.5\\
8&6&0.7\\
9&4&0.7& as Model~7, different sequence of random numbers\\
10&4&0.7& as Model~2, arms jump in position randomly\\
19&4&0.7& small smooth seed field, random injections turned on at $\tau=8$\\
\noalign{\smallskip}
\hline
\end{tabular}
\end{center}
\label{table}
\end{table*}

\subsection{Additional computations}
\label{add}

\begin{figure*}
\begin{center}
\begin{tabular}{lll}
a)\includegraphics[width=0.3\textwidth]{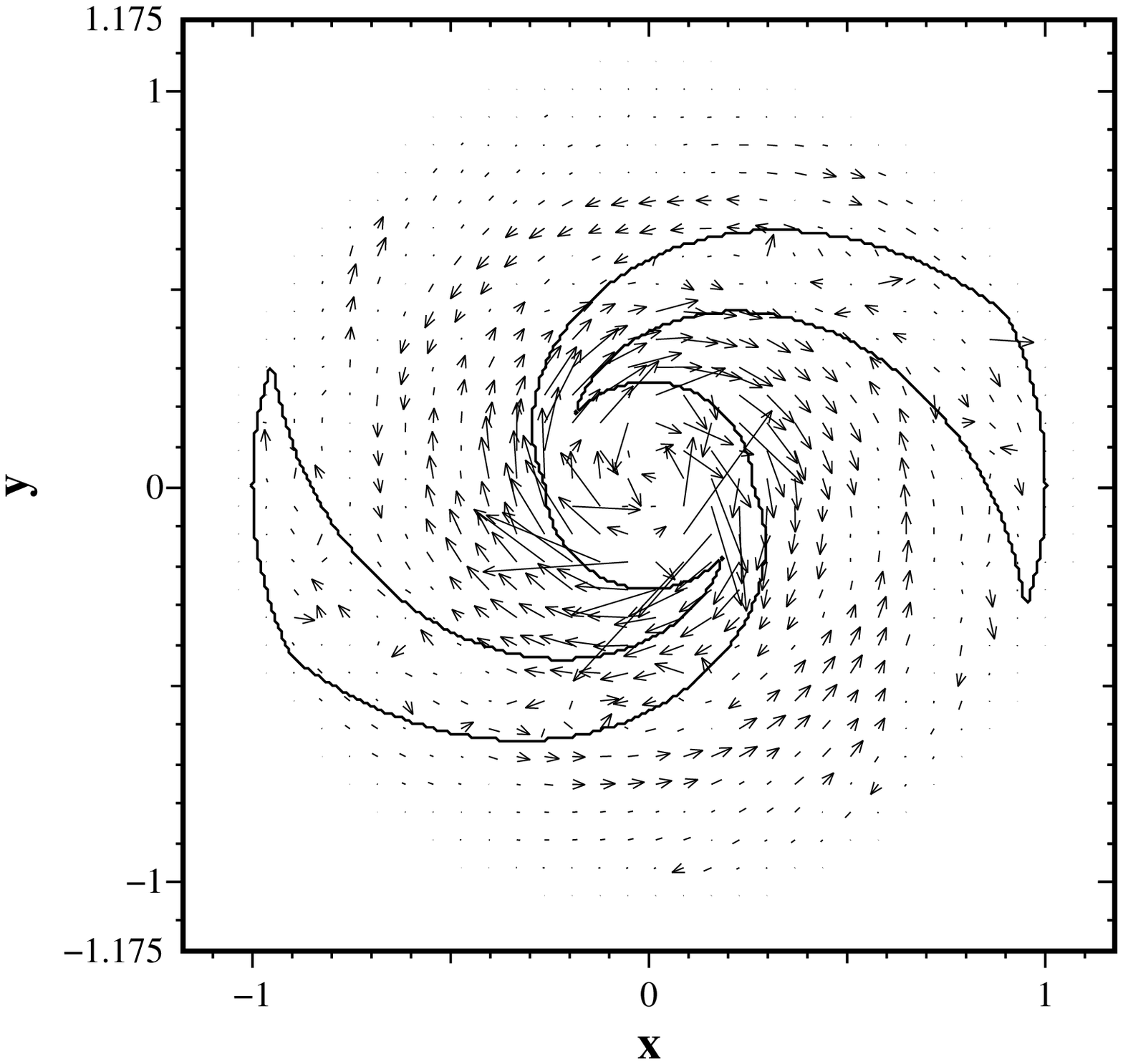} &
b)\includegraphics[width=0.3\textwidth]{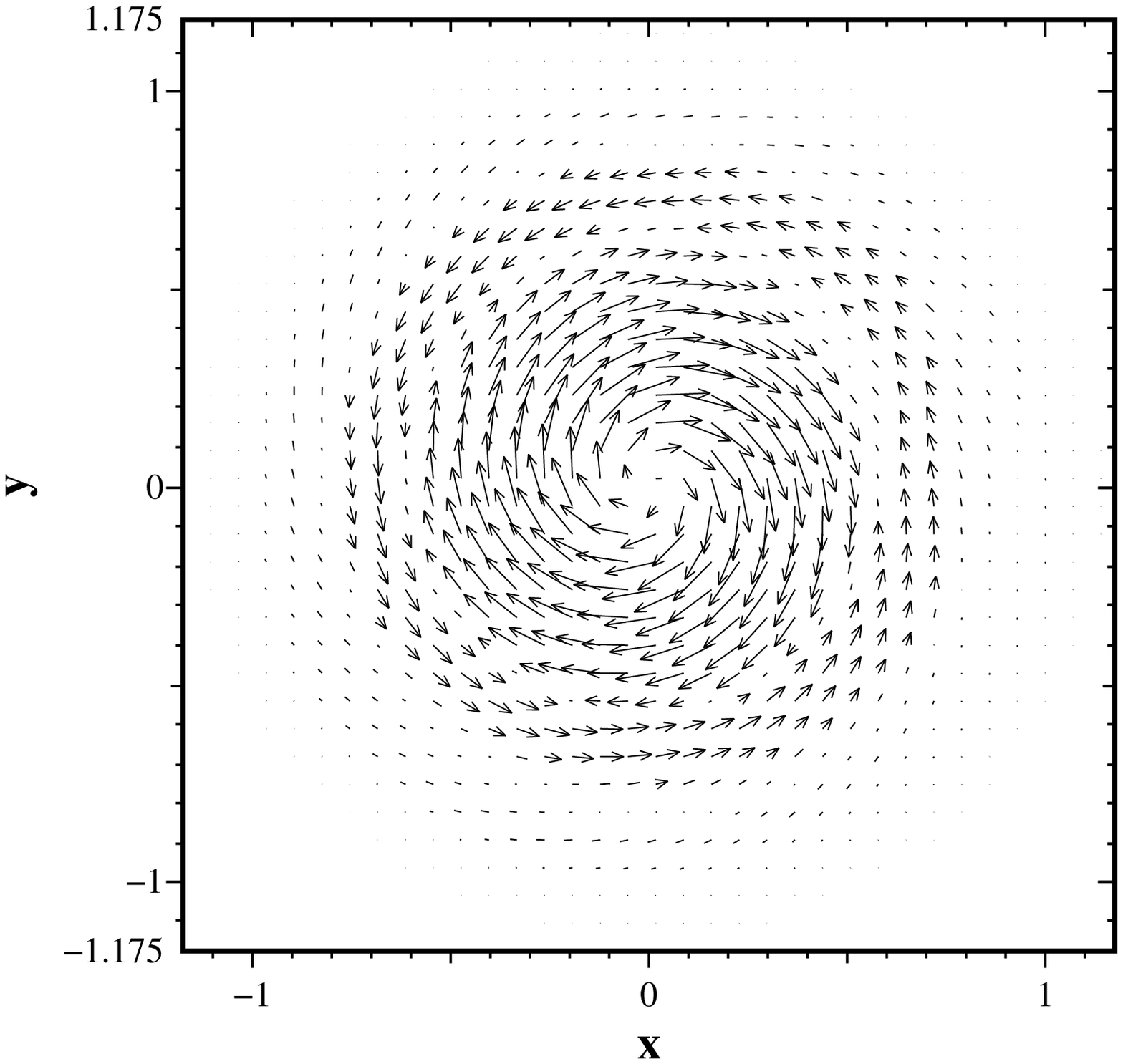} &
c)\includegraphics[width=0.3\textwidth]{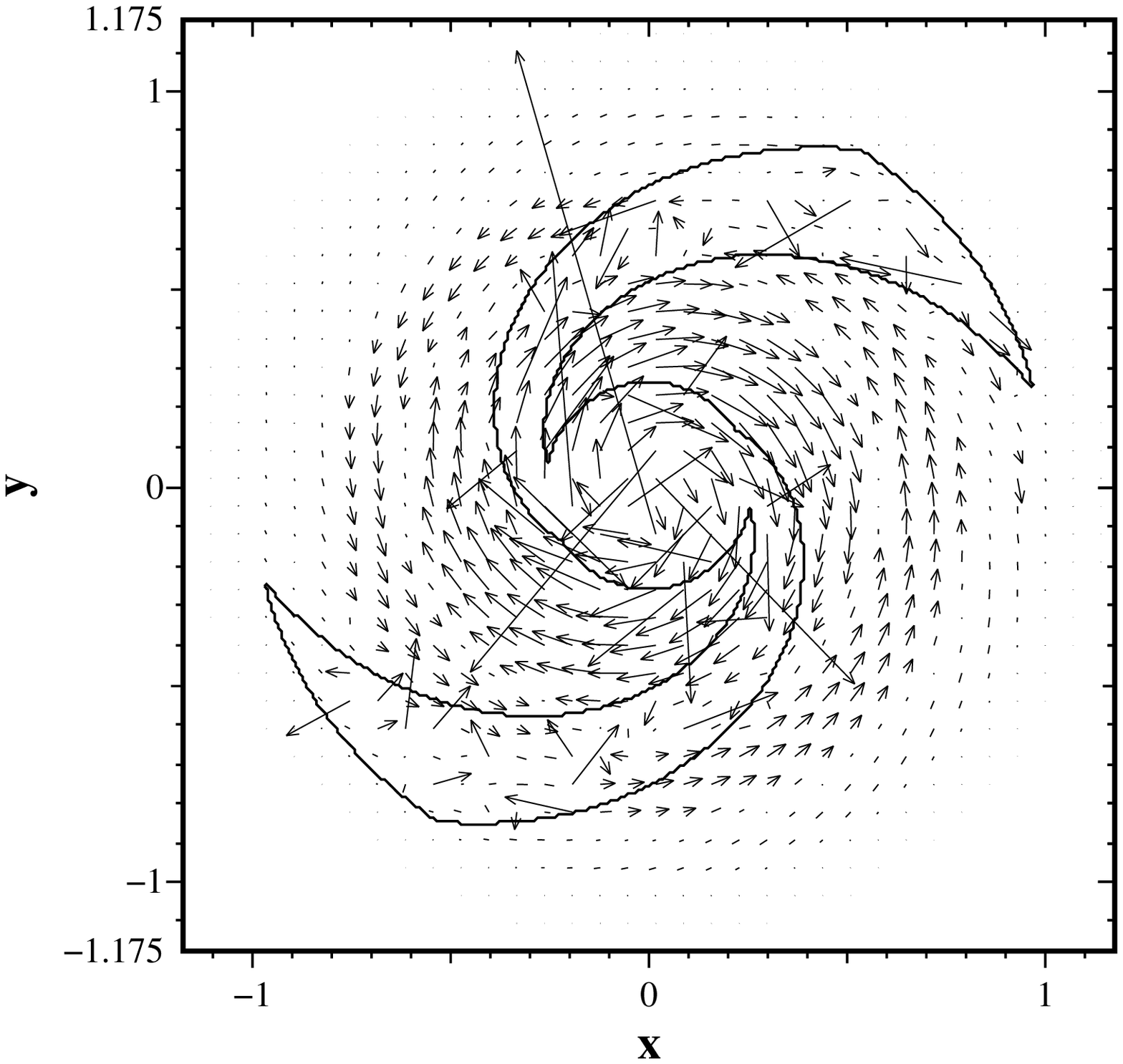} \\
\end{tabular}
\end{center}
\caption{Field vectors at dimensionless times a) $t\approx 8.4$, b)  9.9 and c) 10.6,
for Model~6 with the diffusivity contrast parameter $\eta_{\rm a}=4$.
The arms are removed between $\tau=8.5$ and $10.5$, and panel c) shows
that the interarm regular fields appear almost immediately after the arms reappear.
The anomalously long vectors are the result of a field injection just before $\tau=10.6$.
}
\label{mod186}
\end{figure*}

Simulations strongly suggest that spiral arms are not permanent structures,
but dissolve and reform over comparatively short intervals (see the review by Dobbs \& Baba 2014).
It is also possible
that arms may disappear altogether for intervals and then reappear.
We also conducted two experiments to test how such changes in 
the structure of the material arms might influence our results.
In the first (Model~10), the position of the spiral arms jumps randomly at
time intervals $[t]\approx 5\times 10^8$ yr.
In the second (Model~6 -- see Fig.~\ref{mod186}), the spiral arms are removed altogether for two intervals of about
$1.6 \times 10^9$ Gyr before reappearing.
The conclusion from these computations is that the magnetic interarm structures are present
soon after the material arms reappear.
Model~6 has the same parameters as Model~2 described above, but  the arms
(and associated diffusivity inhomogeneities) are removed for $8.5\le \tau \le 10.5$
and $13.0 \le \tau \le15.0$. Fig.~\ref{mod186} shows field vectors at times
$\tau=8.4$ (i.e. just before the arms are "turned off"), $\tau=9.9$,
 and $\tau=10.6$ (i.e. soon after restoring the arms).
At $\tau=8.4$ field structures are as in Fig.~\ref{mod182}. These disappear almost
immediately the arms are removed, and a near-circular field is
present (Fig.~\ref{mod186}b at $\tau=9.9$). When the arms are turned
back "on" structures similar to those seen in Fig.~\ref{mod182} rapidly
reappear -- see field plots at $\tau=10.6$ in Fig.~\ref{mod186}c.

Additionally, when the arms jump randomly (Model~10), 
similar structures quickly adjust to the new positions of the arms.
Field vectors are not shown for this case.
These idealized experiments suggest that magnetic field structures
are not very dependent on the evolutionary history of the material arms.

\subsection{Roles of small-scale field injections and diffusivity gradients}
\label{roles}

\begin{figure}
\includegraphics[width=0.45\textwidth]{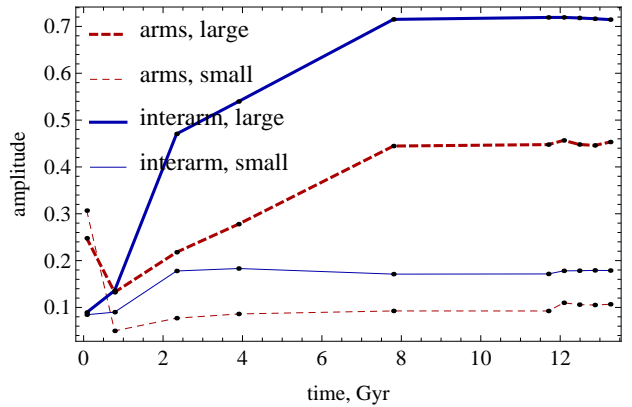}
\caption{Model 7, showing global amplitudes of large- and small-scale field in the arms
(respectively thick and thin red dashed) and large- and small-scale field in the
interarm regions (respectively thick and thin blue curves).}
\label{arm_interarm_187t}
\end{figure}

In order to separate the role of diffusivity gradients alone
in producing the magnetic structures discussed above, Model~7 shows the result
of a computation with parameters as Model~2, except that there are no ongoing field injections. The field rapidly (by $\tau\la 3$) becomes steady. Fig.~\ref{mod187}
shows the field structure (nominally at 13.26 Gyr, i.e, $\tau=17$). Enhanced interarm
fields (magnetic arms) are clearly visible (we note that there is only "regular"
field in this computation).
In Fig.~\ref{arm_interarm_187t} 
the contrast
between the global amplitudes of the regular field in the
arm and interarm regions is clearly visible.

This suggests strongly (and consistently with the suggestion of Moss et al. (2012)
-- and even Moss (1998)) --
that the diffusivity gradients postulated can be largely responsible for the formation
of (regular field) magnetic arms. The small-scale field injections
provide (unsurprisingly) the observed enhanced small-scale field within the
material arms. These injections form a consistent  part of the model in that they
are a direct consequence of the strong star formation in the arms,
 that in turn drives
the turbulence that is responsible for the increased diffusivity there.

\section{Discussion and conclusions}
\label{disc}

We have presented a mechanism for magnetic arm formation based on the 
joint action of turbulent diffusivity contrasts
and small-scale magnetic field injections from small-scale dynamo action,
plausibly associated with supernovae complexes and HII regions, 
although we cannot use this energy input to the ISM directly to calibrate
our parameter $B_{\rm inj0}$. This mechanism produces quite marked large-scale
magnetic structures situated mainly between the material arms
(see, e.g., Fig.~\ref{mod182}). 
However, from time to time a magnetic arm can intersect somewhere a
material arm, e.g. Figs.~\ref{mod182} (right hand panel) and \ref{mod188}. The magnetic arms obtained are quite robust structures and do not require fine tuning of the dynamo governing parameters.

In general, the effect of introducing the diffusivity contrast is
to increase the global mean diffusivity and  to reduce the global mean large-scale
field. However the increased localization of the large-scale field
could result in local enhancements, but in practice this effect seems smaller.
The much increased local diffusivity in the arms results in a more rapid decay
of the injected small-scale field, and so a greater contrast between
large- and small-scale fields, as suggested by the Figures.
A test case omitting the term indicates that
the turbulent diamagnetic velocity $-\frac{1}{2}\nabla\eta$ appearing in 
Eqn.~(\ref{dyneqn}) plays only a minor role.

Our model reproduces the main feature of the effect, but no attempt
has been made to reproduce all details of magnetic arm
formation. In particular, small-scale dynamo action is represented only
 by magnetic field injections. The statistical
properties of injections are obviously simplified, e.g. injections occur
at regular prescribed instants rather than being distributed
more or less homogeneously in time.  Presumably, this statistical inhomogeneity is responsible for  a sharp peak in
evolution of the small-scale field in arms shown in Fig.~\ref{arm-interarm_182t}.
Development of the model in order to obtain a more
realistic description of small-scale dynamo action seems to be an important undertaking.

The pitch angles of the field structures are generally smaller than those of the
spiral arms (similarly to those of Chamandy et al. 2013, obtained using a quite
different mechanism),
but locally have more realistic values, especially in the model (Fig.~\ref{mod188}) with larger diffusivity contrast.
Additionally, their {\bf B}-structures are not aligned along the structures,
 again differing from most observations -- small deviations are
possibly occasionally present. 
In some cases the interarm structures are rather broad compared
to those observed, but  M~81, for example, does have broad interarm structures
(e.g. Krause et al. 1989). Larger diffusivity contrasts than illustrated can give narrower 
and more filamentary arms.
Our model assumes a link between star formation and magnetic arms. Indeed, stronger star formation gives larger
diffusivity contrast which in turn gives more pronounced magnetic arms --
compare Figs.~\ref{mod182} and ~\ref{mod188}. Isolation of a correlation between star
formation and magnetic arms as well as verification of other possible
consequences of the model requires a richer
observational basis for investigation of the magnetic arms.
At the moment it is not very clear how much can be deduced from
observations of several nearby galaxies (NGC~6946 appears likely to yield 
initial results).
Although our models are restricted to the no-$z$ approximation, this does seem
to be quite robust when applied to disc galaxies (e.g. Chamandy et al. 2014).

We recognize that our assumed diffusivity gradients are a theoretical
(but plausible) assumption and are not based directly on observational evidence,
and that there are several other mechanisms that can contribute
to formation of magnetic arms. In particular,
modulation of the alpha effect and time delay between the distributions
of  dynamo drivers and dynamo suppression, the
treatment of helicity fluxes including enhanced outflow in the interarm regions, etc, can all play a role (e.g. Chamandy et al. 2013, 2014, 2015).
We believe however that the mechanism described here,
 based on joint action of the turbulent diffusivity contrast and small-scale
magnetic field injections, gives a natural basis for explaining 
the phenomenon within classical mean-field dynamo theory.

\begin{acknowledgements}RS and DS are grateful to MPIfR for financial support and hospitality during visits to Bonn. DS acknowledges financial support from RFBR under grant 15-02-01407. RB acknowledges support by DFG Research Unit FOR1254.
\end{acknowledgements}

\end{document}